\documentclass[conference,10pt]{IEEEtran}
\IEEEoverridecommandlockouts
\usepackage{cite}
\usepackage{amsmath,amssymb,amsfonts}
\usepackage{algorithmic}
\ifCLASSOPTIONcompsoc
    \usepackage[caption=false, font=normalsize, labelfont=sf, textfont=sf]{subfig}
\else
    \usepackage[caption=false, font=footnotesize]{subfig}
\fi
\usepackage{graphicx}
\usepackage{textcomp}
\usepackage{xcolor}
\def\BibTeX{{\rm B\kern-.05em{\sc i\kern-.025em b}\kern-.08em
    T\kern-.1667em\lower.7ex\hbox{E}\kern-.125emX}}

    \usepackage{ulem}

\IEEEoverridecommandlockouts\IEEEpubid{\makebox[\columnwidth]{ 978-1-6654-3540-6/22/\$31.00~\copyright~2022 IEEE \hfill} \hspace{\columnsep}\makebox[\columnwidth]{ }}
    
\begin{document}

\widowpenalty10000
\clubpenalty10000

\title{Intelligent Blockage Recognition using Cellular mmWave Beamforming Data: Feasibility Study\\
}

\makeatletter
\newcommand{\linebreakand}{%
  \end{@IEEEauthorhalign}
  \hfill\mbox{}\par
  \mbox{}\hfill\begin{@IEEEauthorhalign}
}
\makeatother

\author{

\IEEEauthorblockN{Bram van Berlo\IEEEauthorrefmark{1}, Yang Miao\IEEEauthorrefmark{2}\IEEEauthorrefmark{3}, Rizqi Hersyandika\IEEEauthorrefmark{3}, Nirvana Meratnia\IEEEauthorrefmark{1}, \\
Tanir Ozcelebi\IEEEauthorrefmark{1}, Andre Kokkeler\IEEEauthorrefmark{2}, Sofie Pollin\IEEEauthorrefmark{3}}

\IEEEauthorblockA{\IEEEauthorrefmark{1}Mathematics and Computer Science, Eindhoven University of Technology, the Netherlands}
%\IEEEauthorblockA{\textit{Email: \{b.r.d.v.berlo, t.ozcelebi, n.meratnia\}@tue.nl}}

\IEEEauthorblockA{\IEEEauthorrefmark{2}Radio Systems, University of Twente, the Netherlands}
%\IEEEauthorblockA{\textit{Email: \{y.miao, a.b.j.kokkeler\}@utwente.nl}}

\IEEEauthorblockA{\IEEEauthorrefmark{3}Networked Systems, Katholieke Universiteit Leuven, Belgium}
%\IEEEauthorblockA{\textit{Email: \{rizqi.hersyandika, sofie.pollin\}@kuleuven.be}}

}

\maketitle

\begin{abstract}
Joint Communication and Sensing (JCAS) is envisioned for 6G cellular networks, where sensing the operation environment, especially in presence of humans, is as important as the high-speed wireless connectivity. Sensing, and subsequently recognizing blockage types, is an initial step towards signal blockage avoidance. In this context, we investigate the feasibility of using human motion recognition as a surrogate task for blockage type recognition through a set of hypothesis validation experiments using both qualitative and quantitative analysis (visual inspection and hyperparameter tuning of deep learning (DL) models, respectively). A surrogate task is useful for DL model testing and/or pre-training, thereby requiring a low amount of data to be collected from the eventual JCAS environment. Therefore, we collect and use a small dataset from a 26 GHz cellular multi-user communication device with hybrid beamforming. The data is converted into Doppler Frequency Spectrum (DFS) and used for hypothesis validations. Our research shows that (i) the presence of domain shift between data used for learning and inference requires use of DL models that can successfully handle it, (ii) DFS input data dilution to increase dataset volume should be avoided, (iii) a small volume of input data is not enough for reasonable inference performance, (iv) higher sensing resolution, causing lower sensitivity, should be handled by doing more activities/gestures per frame and lowering sampling rate, and (v) a higher reported sampling rate to STFT during pre-processing may increase performance, but should always be tested on a per learning task basis.
\end{abstract}


\section{Introduction}
%% Yang: JCAS and mmWave human sensing
Millimeter-wave (mmWave) frequencies are used in beyond-5G and 6G mobile communication to provide high-speed wireless connectivity~\cite{6G}. mmWave communication uses directive narrow beams in small cells to countermeasure the propagation loss. However, the mmWave link is easily obstructed by moving humans or objects~\cite{blockage1}. %MmWave communication is much more sensitive to blockage than the sub-6GHz radio communication that is currently used for 4G LTE and Wi-Fi. 
In practice, sensing the operation environment and people in it to predict blockage is required to prevent severe communication performance degradation.

There are several schemes for predicting the human-caused blockage% for assisting the mmWave communication scenarios
, e.g., by using vision/camera~\cite{camera}, radar, LiDAR and sub-6 GHz channel~\cite{sub6}. However, those methods rely on an additional system to predict the blockage, and thus, add complexity. Blockage prediction using in-band mmWave beams was proposed in~\cite{in-bandb}. Pre-blockage signatures, i.e., the received (Rx) signal level fluctuation before blockage happens, was used as input to a Deep Learning (DL) model to predict the potential blockage. The work in \cite{JCS_mmwave_butler} studied target detection by a mono-static 26 GHz platform.

%% Bram: machine learning enabled/assisted human motion recognition
We hypothesize that the signal blockage recognition performance of a DL model can be improved by testing and pre-training it on a surrogate task such as human motion (gesture and activity) recognition using Doppler Frequency Spectrum (DFS) as input. This requires less data collection directly from an operational communication environment. %The difference between gestures and activities can be found both in the Doppler frequency bin region where spectrum magnitude peaks occur, and the peak spread and peak level patterns within that region. 
%The input data structure is different for different motions. %A human torso typically moves much slower than a human arm, thereby showing up in a much lower doppler frequency bin region. 
Activities typically involve movement of more body parts (e.g., legs, torso, and arms) than gestures (hand and fingers), thus showing more peak spread across 
%both low and high 
Doppler frequencies~\cite{1603402}. %10.1145/2971648.2971670}. %This is comparable to a situation where either a small but fast moving bird flies or large and small moving human walks into the field of view of a communication system.

In practice, DFS input data is influenced by the so-called domain factors~\cite{10.1145/3307334.3326081}, which impact the measurements as motions occur. Domain factor examples include operation environment, position of an entity being sensed, and its orientation with respect to transmitter/receiver. %Domain factors influence the way radio signals are transmitted and propagate through space and arrive at a receiver, thereby causing value differences in the input data. 
Varying domain factors results in distribution shifts between data used for training and data encountered during inference (i.e., domain shift), thereby reducing inference performance~\cite{10.5555/3016100.3016186}. %DL models should have the ability to deal with domain factors, up to a certain extent. 
%Capturing large data quantities to reduce risk of domain shift is not always easy. 
Cross-domain features require expert knowledge to define them, rely on assumptions that may not hold in large complex settings involving many domain factors, and may result in %sub-optimal inferencing performance due to 
feature bias~\cite{wigrunt_gu_et_al_2021}.

\begin{figure*}[tb]
\centering
\includegraphics[height=0.32\columnwidth]{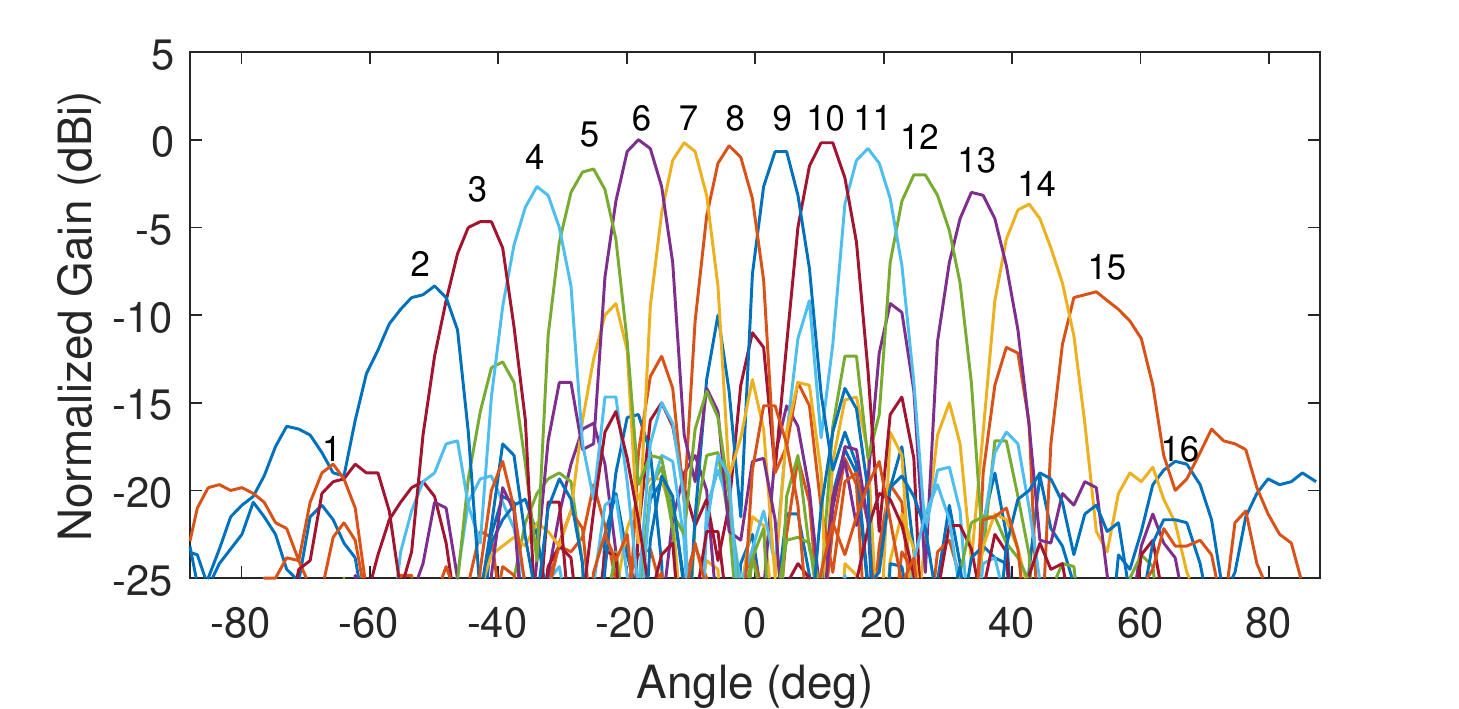}
\includegraphics[width=0.49\columnwidth, height=0.32\columnwidth]{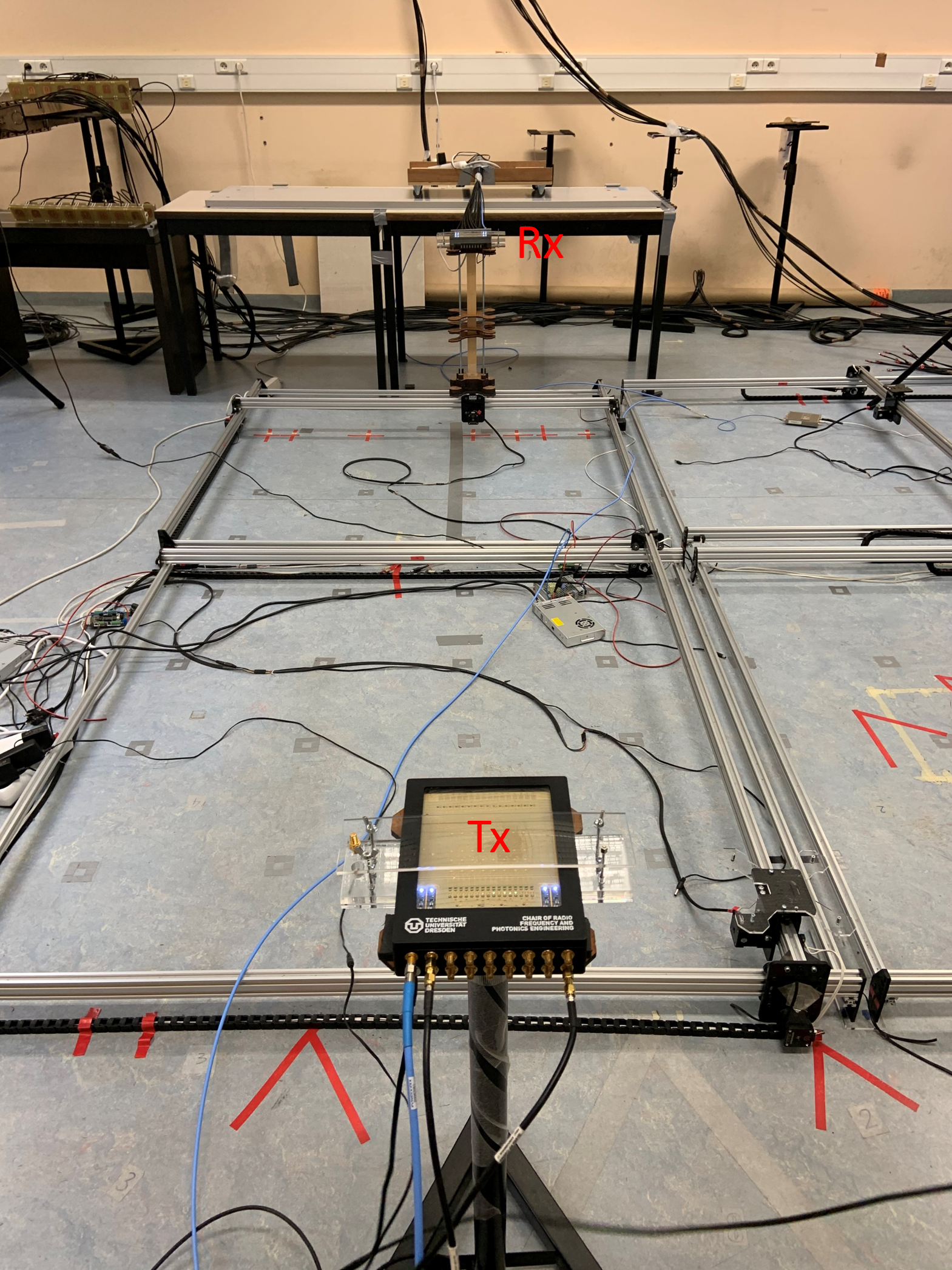}
\includegraphics[width=0.35\columnwidth, height=0.32\columnwidth]{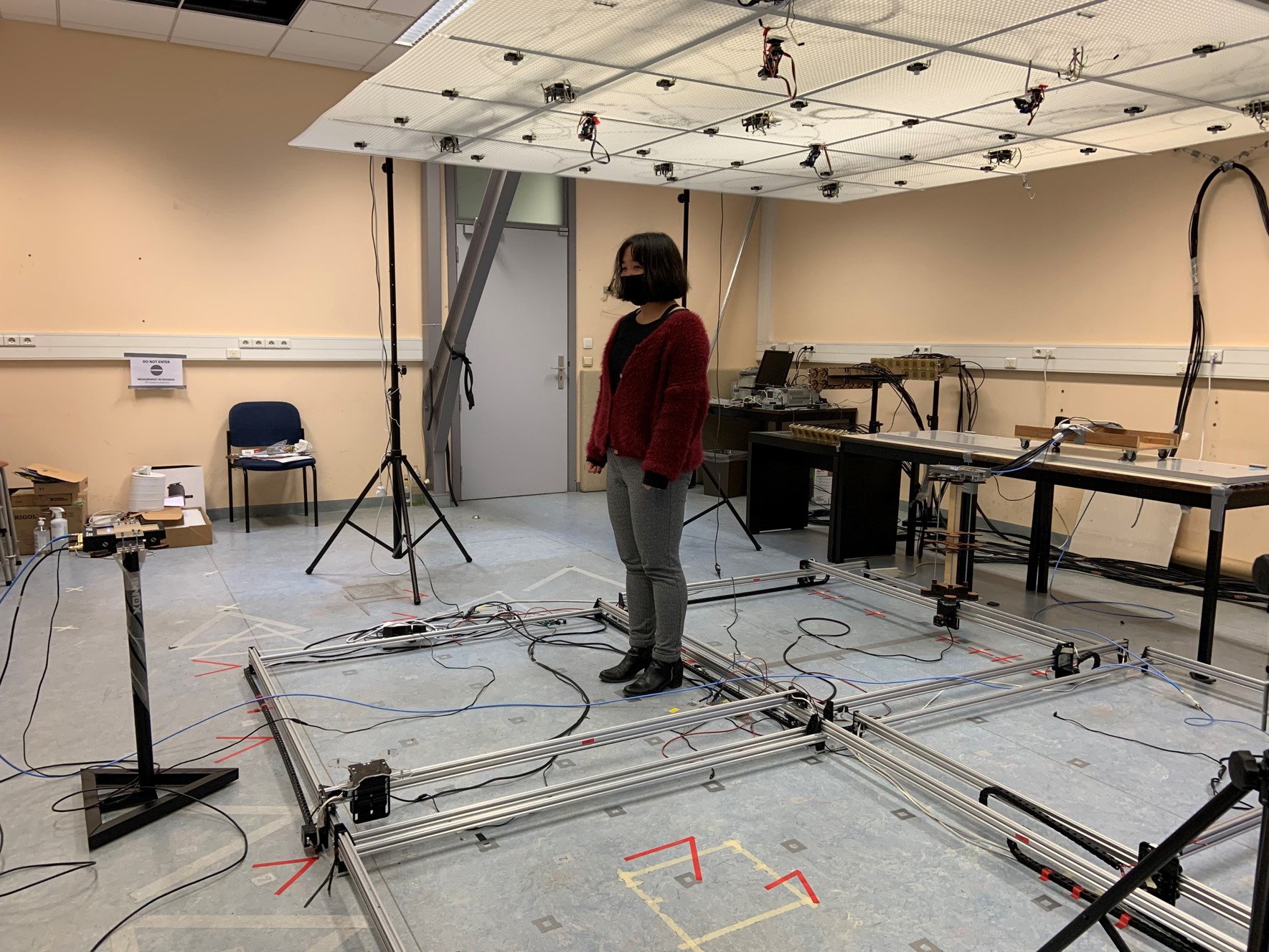}
\includegraphics[width=0.35\columnwidth, height=0.32\columnwidth]{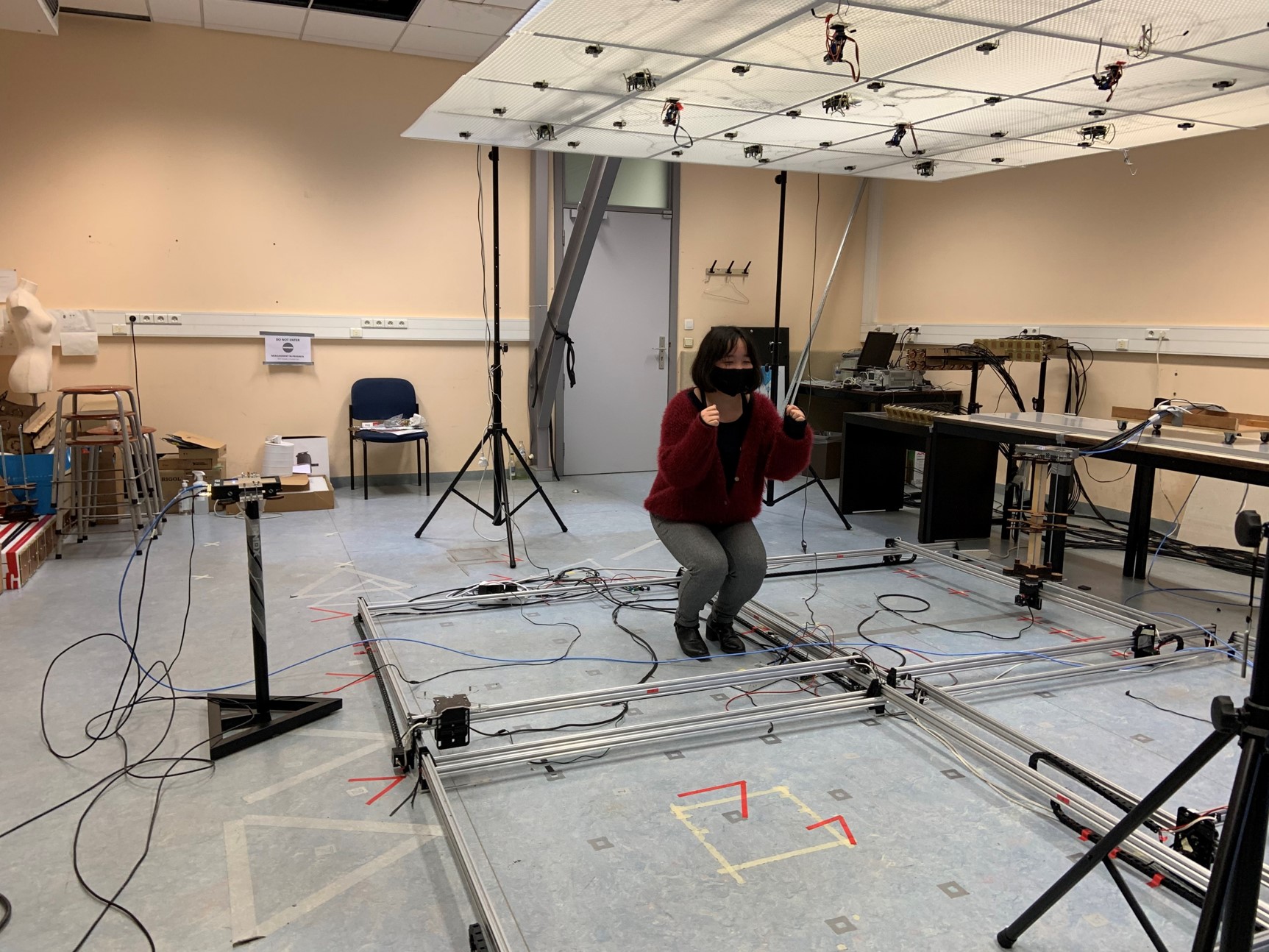}
\caption{From left to right: beam pattern of Butler Matrix at 26 GHz; multi-static BS-UE configuration; target standing still; target doing squats}
\label{figmeasure}
\vspace{-3mm}
\end{figure*}

%% Problem Statement & Motivation of This paper
To the best of our knowledge there is no prior research that uses DFS input data acquired from a mmWave communication device for human motion recognition as a surrogate task for signal blockage type recognition by training DL models with it. We create DFS input data by pre-processing Channel State Information (CSI) matrices acquired in small volume from a 26 GHz cellular multi-user communication device with hybrid beamforming. We consider DL models capable of dealing with domain factors. In this context, the contributions of this paper are in the form of testing the following hypotheses on the input data: 1) Domain shifts exist in the training data which cause inference performance deterioration. %Data samples belonging to the same class, but acquired under different domain factors, contain value differences to such extent that data distribution shifts between samples used during learning and inference are likely to occur in standard DL model data feature representations. This results in relatively reduced inference performance compared to DL models that can handle data distribution shits. 
2) The communication device has hardware sensitivity at a level where one activity or gesture performed per input data DFS frame is enough to distinguish motion class frames from empty class frames. The empty class denotes a situation where nothing is being sensed. This hypothesis assumes that data is acquired under the same domain factors. 3) A small volume of data is enough to obtain reasonable inference performance. %without the need for (i) data augmentation to increase volume or (ii) transfer learning for pre-trained model weight instantiation. 
4) Input data can be split along the receiver antenna dimension to expand the volume of acquired data, thereby leading to increased inference performance across all DL models. 5) Increasing the sampling frequency reported to Short-Time Fourier Transform (STFT) for constructing DFS input samples, thereby changing the reported sampling frequency with respect to the sampling frequency used during CSI matrix collection leads to less aliasing of input data samples, giving better DL model inference performance.
%The contributions are:
%\begin{itemize}
%    \item We introduce human motion recognition as a surrogate task for exploring feasibility of blockage type recognition. The DL models considered have the ability to deal with domain factors. Blockage type recognition serves as initial step in a DL assisted pipeline for blockage avoidance. Based on type, additional steps such as handover can be taken.
%    \item We demonstrate through extensive hypothesis evaluation that, when acquiring CSI matrix data from a 26 GHz communication device and pre-processing it into DFS data, specific data capture and use requirements have to be met. These requirements lead to more successful learning of blockage type recognition and better subsequent blockage type inference performance.
%\end{itemize}

\section{Related Work}
%Domain factors cause input data value differences, especially in mmWave and WiFi-CSI settings. This results in data feature distribution shifts between data used for training and data encountered during inferencing in DL models, thereby deteriorating inference performance of DL models. It is therefore important to consider DL models that can deal with data feature distribution shifts during blockage recognition feasibility exploration. 
Various  DL models for dealing with data feature distribution shifts for human motion recognition, the same surrogate we use for pre-training a model for blockage recognition in this paper, exist. However, to the best of our knowledge, there exists no DL model-based blockage recognition research paper, which considers domain independency or domain shift effects in the field of mmWave and/or WiFi CSI. The DL techniques presented in this section originate from the WiFi-CSI and not the mmWave sensing community because all techniques in the field of mmWave sensing we have seen thus far rely on commercially available mono-static radar hardware setups for data collection. The CSI data extracted from a 26 GHz cellular multi-user communication device with hybrid beamforming (like the setup used by us) resembles data extracted from commercial WiFi-CSI data collection setups. 

Domain adaptation techniques adapt a model trained with data originating from a source domain to a target domain. A `domain' is assumed to consist of an arbitrary fixed number of domain factors uniquely constituting the domain. Yang et al.~\cite{8839094} used two DL model twins, each taking a different input sample from the  source and target domain. The probability distributions over the domains were brought closer together via minimizing an empirical estimation of the maximum mean discrepancy between distributions in a reproducing kernel Hilbert space. Chen et al.~\cite{10.1145/3366423.3380091} used a joint classification-reconstruction task to adapt the representation structure of a small training dataset to a dataset coming from a large pool of unseen domain factors. Soltanaghaei et al.~\cite{10.1145/3408308.3427983} first trained a neural network with data coming from a single source domain. Afterwards, with help of users, the model deployed inside a new home environment was subsequently fine-tuned via transfer learning on newly labeled data.

Domain independent models try to mitigate the effect of domain factors while simultaneously increasing the effect of class differences in feature representations. In the WiFi-CSI sensing community, this is mostly done via steering model training using an adversarial training procedure or with help of an attention mechanism. Most adversarial learning papers~\cite{10.1145/3241539.3241548,10.1145/3380980,9413590,10.1049/ell2.12097,9488693} rely on access to domain labels. During network training, in addition to having a task output that produces class predictions, the network also has a 'discriminator' output that produces domain predictions. The task predictor and discriminator, in an adversarial training procedure, play a min-max game in which the effectiveness of the discriminator is minimized while the effectiveness of the task predictor is maximized. Attention is used to put more cognitive attention towards specific parts of the input sample and to learn the most important correlations between features in the intermediate representation of the DL model. Both~\cite{wigrunt_gu_et_al_2021,9149596} used two attention modules consisting of a sub neural network that learns a function to produce an input mask based on the input given to the sub-neural network. %The attention modules are learned together with the other parts of the DL model.

%% Yang and Rizqi

\begin{figure*}[t]
\vspace{-5mm}
\centering

\subfloat[]{\includegraphics[trim={4.5cm 4.9cm 4.0cm 5.0cm},clip,width=.32\textwidth]{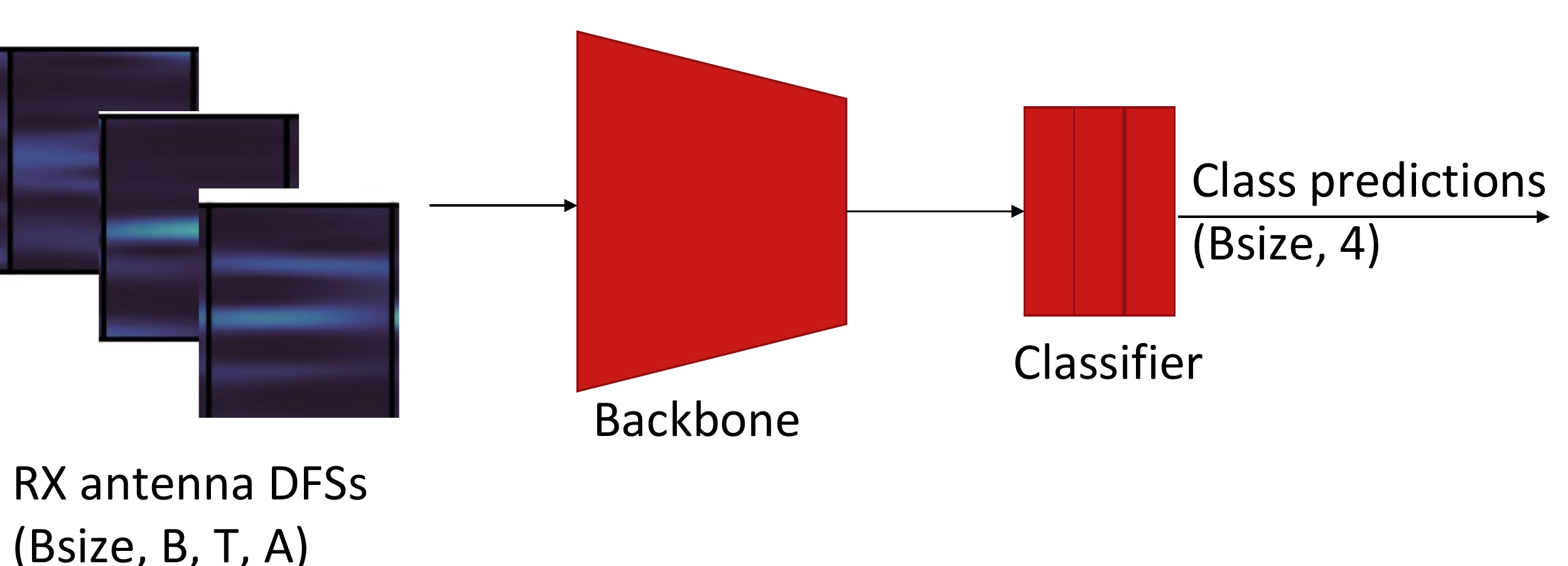}}\hfill%
\subfloat[]{{\includegraphics[trim={4.0cm 2.9cm 1.7cm 1.0cm},clip,width=.32\textwidth]{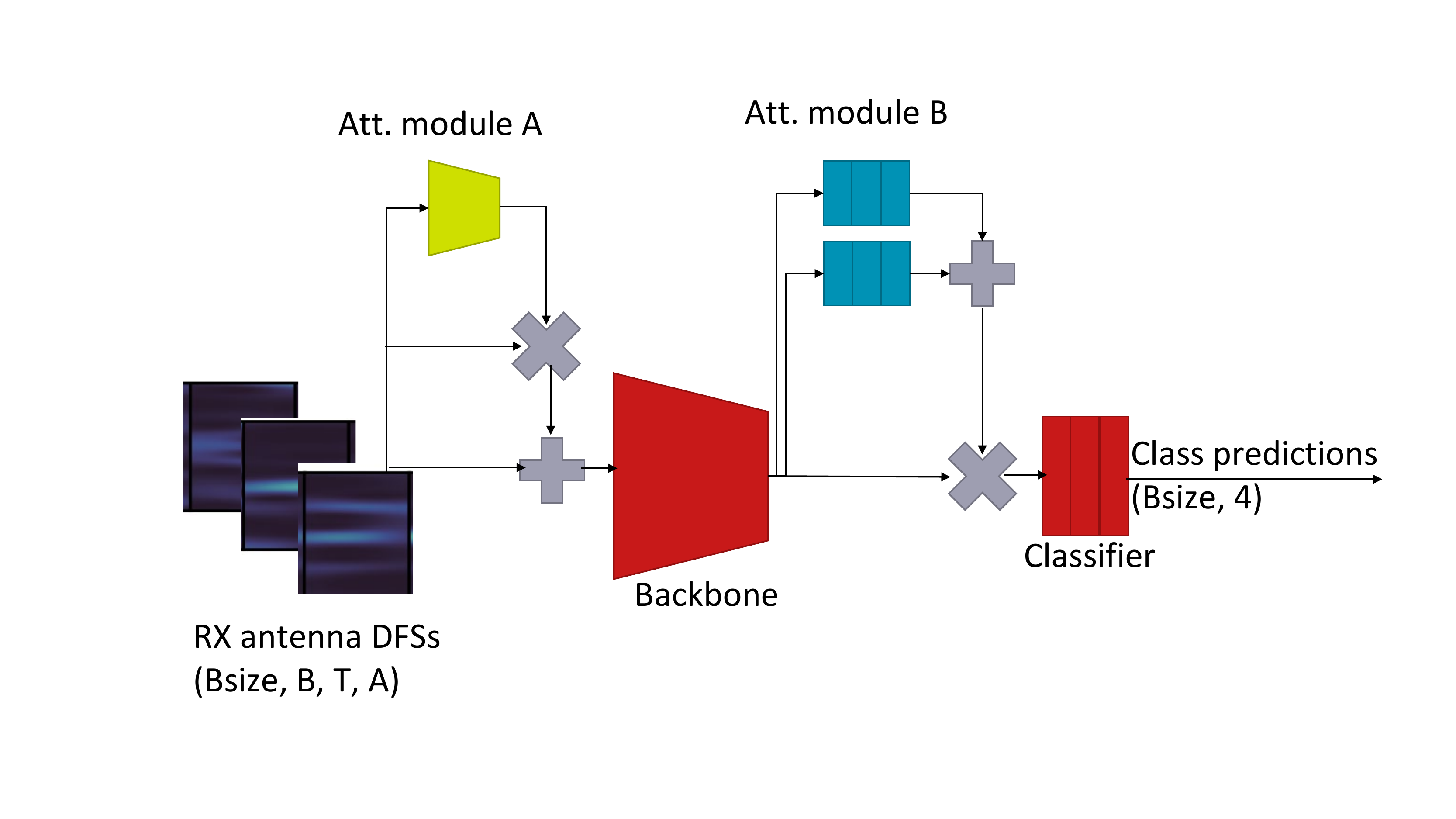}}}\hfill%
\subfloat[]{\includegraphics[trim={5.0cm 2.9cm 4.7cm 1.0cm},clip,width=.32\textwidth]{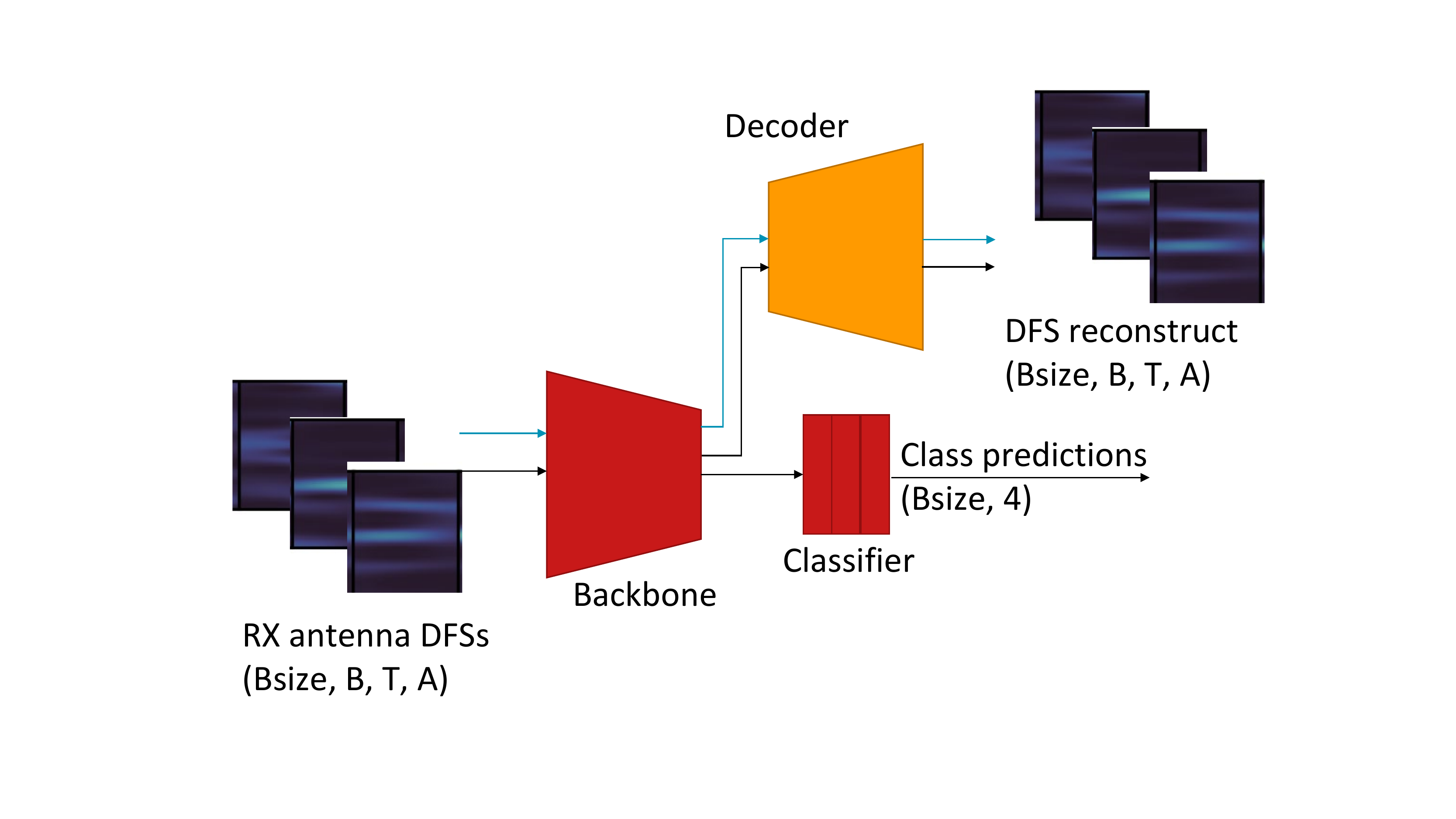}}

%\caption{Models used for hyperparameter tuning. (a) Standard model, which is trained end-to-end using a classification task specific loss which is minimized. (b) Domain generalization model, adopted from~\cite{wigrunt_gu_et_al_2021}, which is supported by two attention submodules, i.e., A prioritizes input features; B correlates features across time originating from the intermediate representation coming out of the backbone. The model is trained end-to-end using a task specific loss which is minimized. (c) Domain adaptation model, adopted from~\cite{10.1145/3366423.3380091}, which involves two data flows: a source (black) and a target (blue) flow. Current input data batch is fed to the source flow followed by feeding a random data batch from the input dataset to the target flow. During training, the domain adaptation model minimizes a task and reconstruction loss sequentially. The reconstruction loss is a summed loss involving a reconstruction loss per data flow. The classification loss i solely based on the source flow.}
%\label{fig:res_meth_dl_models}

\caption{Models used for hyperparameter tuning. (a) Standard model, (b) Domain independency model, adopted from WiGRUNT~\cite{wigrunt_gu_et_al_2021}, (c) Domain adaptation model, adopted from FiDo~\cite{10.1145/3366423.3380091}. These three models are further e4explained in Section \ref{methodology}}
\label{fig:res_meth_dl_models}
\vspace{-3mm}
\end{figure*}

\section{Measurement Campaign}
\label{sec:measurement_campaign}

%\subsection{System Specification}
Our measurement platform consists of two Butler Matrix units and multiple software defined radio USRPs.
The Butler Matrix is a mmWave front-end operating at 26 GHz to 28 GHz frequency ranges. It can transmit or receive 16 orthogonal beams. The patterns of the beams are frequency-dependent and those at 26 GHz are shown in Figure~\ref{figmeasure}.
The beams are more focused at the center and are less focused at the edge, causing up-to-20 dB gain difference. Due to this characteristic, we only use beam 3-14 at the transmitter.     
The USRPs transmit signals with intermediate frequency of 2.4 GHz and 20 MHz bandwidth to the transmitter (Tx) Butler Matrix, while the Local Oscillator (LO) port of the Tx Butler Matrix is connected to the ERASynth+ RF Signal Generator operating at 11.8 GHz to up-convert and generate signals at 26 GHz. The transmit power is 0.1 mW. At the Rx, the LO port is connected to the frequency synthesizer synchronized for Tx and Rx. Additional Rx gain of 31.5 dB is added to the Rx USRPs. The Tx and Rx USRPs are synchronized using Pulse Per Second (PPS) and 10 MHz reference signals, via the National Instrument LabView Communications MIMO Application Framework. The framework uses a Time Division Duplex (TDD) signal frame structure with Orthogonal Frequency Division Multiplexing (OFDM) symbols. The channel estimation is based on the uplink pilot. The noise floor is about -93.85 dBm.

%\subsection{Measurement Scenario}
The measurements were conducted in a room with size of $11\times6.5$ m$^2$. As shown in Figure~\ref{figmeasure}, the multi-static Tx-Rx configuration was used to mimic the BS-UE channel.
%, while the other was to mimic scenario of using the monostatic BS Tx-Rx functioning as radar for contact-free sensing. 
The heights of the Tx/Rx Butler matrices were located 90 cm above the ground. When the Tx and Rx beams are aligned, the channel has the maximum gain.
%As shown in Figure~\ref{fig2}, the sensing target, a female with 161 cm height and medium-fat body type, was standing inside the first Fresnel zone of the field of view (FoV) of all beams, was doing squats (3 squats in 5 seconds), or was standing with twirling fingers. 
%For communication, we used multi-static BS-UE configuration where the Tx and Rx beams must be aligned to achieve maximum gain.
%
The measured radio channel at one time was $16 \times 1$ with 16 receiving beams and 1 transmitting beam. 
We did the same measurement for multiple times with different $n_{tx}$ Tx beam to obtain the channel with dimension of $16 \times n_{tx}$, assuming the reproducible target movement. 
A time symbol duration inside a measurement contains 1600 complex channel samples from 16 receiving antennas and 100 subcarriers.
%Rizqi
%\subsection{Data Conversion}
The CSI data contain the complex channel state in 3 dimensions: Rx antenna, subcarrier (frequency), and symbol (time). It was recorded in the binary (.bin) format before being converted to .csv format for further processing.

%% Bram
%\section{Research Methodology}
\section{Blockage Recognition Pipeline}
\label{methodology}
For our hypothesis validation, we perform two experiments: (i) visual inspection of CSI matrices converted to DFS plotted along an image grid; (ii) hyperparameter tuning. 

%In what follows, we explain our methodology. 
For data pre-processing,
%\subsection{Data Pre-processing}
%\label{sec:data_preprocessing}
we first convert the acquired CSI matrices to DFS input data. Data pre-processing starts with ordering and stacking CSI matrices loaded from CSV files according to a tensor \(\in \mathbb{R}^{(A\times K\times T)}\), unless stated otherwise. Symbol $A$ denotes receiver antenna beams, $K$ sub-carriers, and $T$ a certain time period. The tensor is subsequently dimensionality reduced in the sub-carrier dimension to the first principal component using Principal Component Analysis (PCA). The resulting tensor is transformed into DFS input tensor structure \(\in \mathbb{R}^{A\times B\times T}\) using STFT across the time dimension per receiver antenna beam. Symbol $B$ denotes Doppler frequency bins. Lastly, the Doppler frequency bin and time dimensions are zero padded to the nearest power of 2, increasing neural network depth search interval during hyperparameter tuning.

\vspace{-1mm}
\subsection{Hyperparameter Tuning} %With Domain Adaptation/Generalization}
\label{sec:model_task_details}
%Hyperparameters are those model parameters that in addition to trainable model parameters, i.e. weights and biases, should be defined prior to model training. Optimal hyperparameters are often found from a high-dimensional search space in case these hyperparameters are unknown based on previous research. We explore the search space by consecutively running a vast number of DL model training procedures based on the hyperband algorithm.

Our hyperparameter tuning based on hyperband~\cite{10.5555/3122009.3242042} involves three models: a standard model, a domain adaptation model, and a domain independent model, shown in Figure~\ref{fig:res_meth_dl_models}. When these models are trained, input data is fed into the models using minibatch tensors \(\in \mathbb{R}^{Bsize\times B\times T\times A}\). %Model task, architecture details, and the hyperparameter tuning concept are further explained in the following subsection. %Subsections~\ref{sec:model_task_details} and~\ref{sec:model_arch_details}.  %The reason why no further cross-validation experiment is considered is that it does not contribute significantly more insights for hypothesis validation in comparison to the hyperparameter tuning experiment.
The standard model consists of two components: a backbone network and a classifier network. The purpose of the backbone network is to create intermediate representations of the input data. %Simple representations include color edges and more complex representations magnitude peak activity sections for example. 
The purpose of the classifier network is to map this to a categorical distribution \(\{P(y_i|z)\}_{i=1}^N\) per sample in the minibatch tensor. Symbol \(y_i\) denotes a specific motion class, \(z\) an intermediate representation of DFS input sample, and \(N\) the total number of motion classes considered. %During learning, the difference between categorical distribution created by the DL model and one-hot encoded class label created by a human labeler is minimized using gradient descent and a particular loss function. Partial loss derivatives w.r.t. a learnable model parameter are computed using the chain rule and used to 
Model parameters are updated according to 
%\eqref{eq:learn_model_param_update} 
%\begin{align}
%\label{eq:learn_model_param_update}
    $W^{(i)}_n = W^{(i)}_n - \eta\frac{\partial L}{\partial W^{(i)}_n}$,
%\end{align}
where \(W\) denotes a weight matrix, \(i\) layer superscript, \(n\) layer parameter subscript, \(L\) loss function, and \(\eta\) the learning rate. %Additional information on deep learning can be found in~\cite{Goodfellow-et-al-2016}.

The domain independent model is the same except that it has extra components: attention modules A and B~\cite{wigrunt_gu_et_al_2021}. The purpose of module A is to create a single input sample representation, which when combined with the original input sample, mitigates irrelevant input sample parts, thereby optimizes the overall loss minimization process further. The purpose of module B is to find important correlations between intermediate representation features across time and put more emphasis on these correlations in subsequent neural network components. The attention modules are simultaneously trained with the other neural network components end-to-end. 
%How the attention modules are chained together with the other neural network components determines the way partial loss derivatives are computed for attention model parameters. More explanation on attention module is in ~\cite{wigrunt_gu_et_al_2021}.

The domain adaptation model has a decoder component with source and target flows. Flows are not the same as input modalities, as each uses the same single input modality propagated sequentially. The current input DFS sample is given as the source flow and a randomly picked input DFS sample is given as the target flow, to the model. Chen et al.~\cite{10.1145/3366423.3380091} explain that unlabeled input samples coming from a much larger target dataset involving many more domain factors are normally considered for the target flow. However, as our dataset is small, this target data requirement cannot be met, whose effects are shown in Section~\ref{sec:results}. In a single training run using a source and target sample, a classification loss is computed based on the categorical distribution created by the DL model and a one-hot encoded class label created by a human labeler of the source flow. The decoder %component, using an inverse version of the backbone component, 
reconstructs the input samples. % belonging to the source and target flows. 
The sum of reconstruction losses for both flows gives the overall reconstruction loss. Partial loss derivatives are computed for both classification and overall reconstruction losses and used to update the trainable model parameters separately in a sequence (see \cite{10.1145/3366423.3380091} for more details).

%\subsection{Neural Network Architectures}
%\label{sec:model_arch_details}
\subsection{Neural Network Architecture}
Our backbone network, which is common in all three models as shown in Figure~\ref{fig:res_meth_dl_models},  starts with a normal 2D convolution block consisting of a convolution layer, swish activation function, and max pooling layer. The convolution filter number search interval is \(A\)-24. The convolution kernel size search interval is 12-48,12-48. Convolution stride is 1,1. The convolution layer does not include bias addition. The max pool size search interval is 2-48,2-48. Max pooling stride is 2,2. The max pooling and convolution layers both apply padding to keep in-/output sizes the same. The convolution block is followed by a variable number of mobile inverted bottleneck~\cite{8578572} and squeeze-an-excitation blocks~\cite{8701503} depending on a value from the backbone depth search interval 2-5. Each block, excluding subsequent block repetitions, is followed by a max pooling layer. The input channel size is dependent on the output channel size of the previous layer and the output channel size is twice the input channel size. Every block uses expansion rate search interval 2-8, a binary decision to include a residual connection, stride of 1,1, squeeze-and-excitation rate search interval 0.05-0.7, does not include dropout, and includes layer repetition search interval 0-2. ReLU6 activation functions are switched out for swish activation functions. The pool size search interval is 2-48,2-48. Max pooling stride is 2,2. Max pooling applies padding to keep in-/output sizes the same. The mobile inverted bottleneck and squeeze-and-excitation blocks are followed by a  2D convolution block consisting of a convolution layer, swish activation function, and max pooling layer. The convolution filter number is twice the output channel size of the last inverted bottleneck and squeeze-and-excitation block. The convolution kernel size search interval is 4-16,4-16. Convolution stride is 1,1. The convolution layer doesn't include bias addition. The convolution layer applies padding to keep in-/output sizes the same. Global max pooling is added to create the latent representation.

Our classifier is a Multi-Layered Perceptron (MLP) firstly made up of a variable number of dense layers depending on a value from classifier depth search interval 1-6. The neuron search interval per layer is 16-1024. Every layer uses a ReLU action function. After the variable number of dense layers, we add another dense layer, which has the number of task classes as number of neurons and uses the softmax activation function to create a categorical distribution. Weights and biases are initialized via random numbers drawn from a variance scaled uniform distribution with scale factor $\frac{1}{3}$ and `fan\_out' mode.

Attention module A is a normal 2D convolution block consisting of a batch normalization layer followed by a 2D convolution layer and sigmoid activation function. The convolution filter number is 1. The convolution kernel size search interval is 16-88,16-88. Convolution stride is 1,1. The convolution layer does not include bias addition and applies padding to keep in-/output sizes the same. Prior to passing an input representation to the backbone network, the attention module A output is element-wise multiplied with the DFS input. The result is element-wise added together with DFS inputs and passed on to the backbone network.

Attention module B starts by forking the intermediate representation into two and applying average and max pooling to either of the forks. Both average and max pooling consider a pool size search interval of 8-48,8-48, stride search interval of 16-32,16-32, and apply padding to keep in-/output sizes the same. Subsequently, each fork is passed in sequence into a shared MLP. The MLP flattens the input to a single dimension. The flat output is then passed on to a variable number of dense layers depending on an attention module B depth search interval 1-6. The neuron search interval per layer is 16-2048. After the variable number of dense layers, another dense layer is added which has the original flat output size given as input size to the variable number of dense layers. Every dense layer uses a ReLU activation function. The output of the dense layers is reshaped into the original input shape prior to passing the forks to the shared MLP. The forks are then element-wise added, passed through a sigmoid activation function, and element-wise multiplied with the intermediate representation coming out of the backbone network.

The decoder network itself is directly connected to the output of the mobile inverted bottleneck and squeeze-and-excitation blocks in the backbone network. Most of the decoder is an exact reverse copy of the mobile inverted bottleneck and squeeze-and-excitation blocks in the backbone network. However, max pooling layers are taken out of the decoder network and replaced with two dimensional upsampling layers. The reversed mobile inverted bottleneck and squeeze-and-excitation blocks are followed by two normal convolution blocks consisting of a convolution layer and swish activation function. The first convolution block also includes a subsequent upsampling layer. Both convolution layers have a kernel size search space of 12-48,12-48, stride of 1,1, do not include bias addition, and apply padding to keep in-/output sizes the same. The first convolution layer has a filter number equal to the filter number of the first convolution layer in the backbone network. The second convolution layer has a filter number equal to the DFS input channel size. Each upsampling layer uses a repetition size 2,2 and nearest value interpolation.

Before starting a training run, an overall decision can be made to include batch normalization layers into most architecture parts. These layers are added to every block mentioned above as first layer, except for the classifier network, squeeze-and-excitation part of every mobile inverted bottleneck block, and both attention modules. Attention module A always includes a batch normalization layer. All convolution layer weights are initialized randomly by drawing from a variance scaled normal distribution with scale factor 2, `fan\_out' mode.

\begin{figure*}[t]
    \centering
    \includegraphics[width=\textwidth]{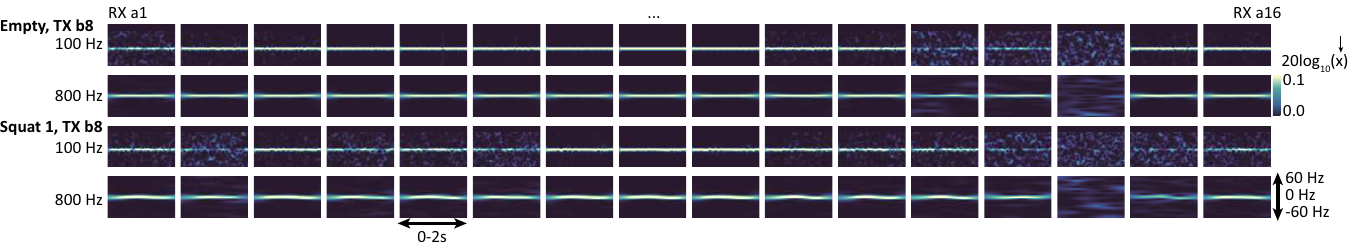}
    \caption{Visual inspection of DFS input sample from dataset 2. For two different acivity classes and one TX beam, two different reported sample frequencies to the STFT pre-processing procedure for samples per RX antenna are shown. The first purpose is to analyze degree of variation against samples from dataset 1. The second purpose is to analyze the effect of changing the reported sampling frequency.}
    \label{fig:vis_inspec_dset2_txb8_rep_sample_rate_comparison}
    \vspace{-2mm}
\end{figure*}

\section{Hypothesis Validation Experiments and Results}
\label{sec:results}

\subsection{Datasets}
%During the measurement campaign detailed in Section~\ref{sec:measurement_campaign}, two different datasets were collected. %We refer to these datasets in Section~\ref{sec:results} as dataset 1 and 2 respectively. 
We use two datasets. Dataset 1 contains data for all Rx antenna patches, Tx beams 5-10 (see Figure~\ref{figmeasure}), and 1 female test subject. TX beams 2-4 and 11-16 are not considered because zero amplitude and phase values were noted when pre-processing CSI matrices. The communication device zeroes signals with a too low Signal-to-Noise Ratio (SNR). The female test subject was asked to perform a single sample repetition per Tx beam of the following 3 motion types: standing still in frontal facing direction, squatting in frontal facing direction, and squatting in orthogonal facing direction. During a sample repetition, the motion was performed 3-4 times. A single repetition of the empty room per Tx beam, denoting the empty class, was also added to the dataset, making the total number of motion recognition classes 4. The 24 DFS input data samples were unstacked (i.e., diluted) in the Rx antenna patch dimension, thereby making the total number of samples in the dataset 24 \(\cdot\) 16 = 384. Total number of unique domains considered is 16 RX antenna patches \(\cdot\) 6 Tx beams = 96. The sampling rate used during data collection was 20Hz. During pre-processing, a sampling rate of 500Hz was given to the STFT procedure. Because the female subject performed the motion recognition activities in a period of 5 seconds, the total number of indices in the time dimension is 100.

Dataset 2 contains data for all Rx antenna patches, Tx beams 7-9, frontal and orthogonal facing direction, and 3 male test subjects. Three male test subjects were asked to perform 3 repetitions per Tx beam and facing direction of the following 3 motion types: standing still, standing still and performing fine-grained hand movement, and squatting. During each repetition, the motion was performed one time. Three repetitions of the empty room per Tx beam, denoting the empty class, were also added to the dataset per male test subject, making the total number of motion recognition classes 4. Additionally, a fourth male test subject joined the data campaign, for Tx beam 7 and frontal facing direction, 3 repetitions of the standing still and performing fine-grained hand movement class. Three repetitions of the empty class at Tx beam 7 were added as well for this subject. The sampling rate used during data collection was 100Hz. Because the male subjects performed the motions in a period of 2 seconds, the total number of indices in the time dimension is 200.
For hypothesis validation experiments, we use three versions of dataset 2:
\begin{enumerate}
    \item DFS input samples were not unstacked. The total number of samples is (3 \(\cdot\) 2 \(\cdot\) 3 \(\cdot\) 3 + 3 \(\cdot\) 3) \(\cdot\) 3 + 2 \(\cdot\) 3 = 195. Total number of unique domains is 2 \(\cdot\) 3 \(\cdot\) 4 = 24. During pre-processing, a sampling rate of 100Hz was given to the STFT procedure.
    
    \item DFS input samples were unstacked in the Rx antenna patch dimension. Therefore, the total number of samples is ((3 \(\cdot\) 2 \(\cdot\) 3 \(\cdot\) 3 + 3 \(\cdot\) 3) \(\cdot\) 3 + 2 \(\cdot\) 3) \(\cdot\) 16 = 3120. Total number of unique domains is 2 \(\cdot\) 3 \(\cdot\) 4 \(\cdot\) 16 = 384. During pre-processing, a sampling rate of 100Hz was given to the STFT procedure.
    
    \item Same as version 2 but during pre-processing, a sampling rate of 800Hz was given to the STFT procedure
\end{enumerate}
% In the third version, DFS input samples were unstacked in the RX antenna patch dimension. Therefore, the total number of samples is ((3 \(\cdot\) 2 \(\cdot\) 3 \(\cdot\) 3 + 3 \(\cdot\) 3) \(\cdot\) 3 + 2 \(\cdot\) 3) \(\cdot\) 16 = 3120. Total number of unique domains is 2 \(\cdot\) 3 \(\cdot\) 4 \(\cdot\) 16 = 384. During pre-processing, a sampling rate of 800 Hz was reported to the STFT procedure. 

\vspace{-2mm}
\subsection{Assessment Strategy}
%The standard, domain adaptation, and domain generalization models are trained by means of cost minimization.
The cost minimization objective for the standard model, domain independent model, and the classification portion of the domain adaptation model is given by
%in Equation~\ref{eq:std_gen_adap_class_cost}. 
%\begin{align}
%\label{eq:std_gen_adap_class_cost}
    $\min_{\theta} - \frac{1}{m}\sum_{i=1}^{m}\sum_{j=1}^{n_l} y_{i, j} \ln \hat{y}_{i, j}$.
%\end{align} \vspace{-4mm}
Symbol \(\theta\) denotes the model parameters, \(m\) the minibatch size, \(n_l\) the number of motion recognition labels, \(y_{i, j}\) the true motion recognition class label value (0 or 1), and \(\hat{y}_{i, j}\) the predicted motion recognition class label value (0-1). The cost minimization objective of the reconstruction portion of the domain adaptation model is given by
%\begin{align}
%\label{eq:adap_reconstr_cost}
    $\min_{\theta} \sum_{t \in \{t_{source}, t_{target}\}} \frac{1}{m_t} \sum_{i=1}^{m_t} (x_{t, i} - \hat{x}_{t, i})^2$.
%\end{align}
%in Equation~\ref{eq:adap_reconstr_cost}. 
Symbol \(t\) denotes a task placeholder for both the source and target reconstruction tasks. %It also indicates that symbols are task specific and thus not shared between tasks. 
The symbol \(x_{t, i}\) is an input data sample and \(\hat{x}_{t, i}\) the reconstructed version of that sample.

During training, gradient descent was performed by means of Adaptive Moment Estimation (ADAM). %For the classification and reconstruction task in the domain adaptation model, two different optimizer objects were used. 
The learning rate, batch size, and maximum number of epochs considered were 0.0001, 12, and 100. %The learning rate, amount of epochs and batch size were determined with manual hyperparameter tuning. 
The datasets considered for training were firstly split into 69\% training, 14\% validation and 17\% test subsets.% based on splitting domain labels associated with the datasets. The train subset was further split into a 69\% train and 14\% validation. 
On the domain labels associated with the datasets, a random 6-fold cross-validation split procedure was performed. The same procedure was subsequently performed on the training subset. To allow better reproducibility of the results reported in this section, pseudo-randomness was introduced to the splitting procedure by adding a seed of 42. The reason no domain-leave-out cross-validation was considered is that overfitting, due to the small data volume present inside the datasets, overshadows the domain shift effect and is therefore inappropriate for the hyperparameter tuning experiment.

\subsection{Qualitative: Visual Inspection}
\label{sec:vis_data_inpsec}
Comparing the sample per Rx antenna in a vertical direction between Tx beams 5 and 9 separately per motion recognition class in Figure~\ref{fig:vis_inspec_dset1_txb_comparison}, we see structural differences among RX antennas. Comparing the sample per Rx antenna for Tx beam 9 in a vertical direction between the different motion recognition classes in Figure~\ref{fig:vis_inspec_dset1_txb9_activity_comparison}, we also observe structural differences in specific Rx antennas. This means that a DL model should be able to distinguish between classes. However, %what can also be noted is that 
many images look exactly the same. Comparing figures~\ref{fig:vis_inspec_dset1_txb9_activity_comparison} and~\ref{fig:vis_inspec_dset2_txb8_rep_sample_rate_comparison}, we see that input samples in dataset 2 contain less variation. This is caused by the instruction given to the male subjects to perform less motions per sample repetition.

\begin{figure}[h]
\vspace{-2mm}
    \centering
    \includegraphics[width=\columnwidth]{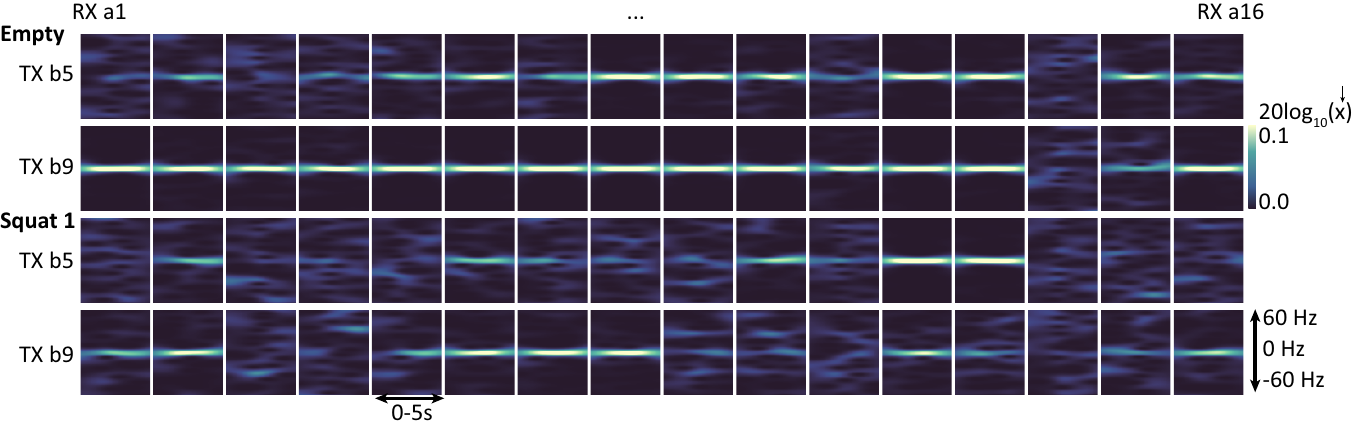}
    \caption{Visual inspection of DFS input sample from dataset 1. For two different activity classes, DFS samples per RX antenna across two different TX beams are shown to analyze potential TX beam factor differences.}
    \label{fig:vis_inspec_dset1_txb_comparison}
\end{figure}

\begin{figure}[h]
\vspace{-3mm}
    \centering
    \includegraphics[width=\columnwidth]{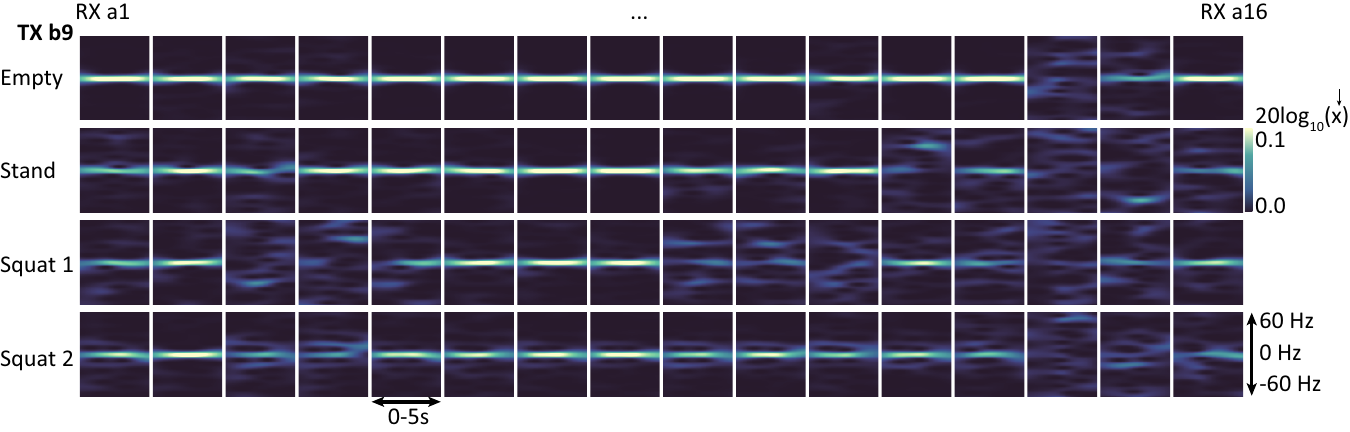}
    \caption{Visual inspection of DFS input sample from dataset 1. For one TX beam, four different activity classes across samples per Rx antenna are shown to analyze the degree of structural differences between classes.}
    \label{fig:vis_inspec_dset1_txb9_activity_comparison}
\end{figure}

\subsection{Quantitative: Hyperparameter Tuning}
\label{sec:hyperparam_tuning}
%We use hyperband algorithm for our hyperparameter tuning experiment. 
During hyperband tuning, an epoch budget of 1000 per iteration, 5 iterations, and a model discard proportion per iteration of 3 were used. Hyperband tuning started with 1000 model candidates randomly sampled from the high-dimensional search space, which were all trained for 2 epochs. Every iteration, the 1/3 best performing models were kept with respect to the previous iteration and trained for more epochs until a low number of well performing model candidates remain. To optimize the hyperparameter tuning experiment, including the training procedures executed during this experiment, only the first fold in the cross-validation splitting procedure was considered during hyperparameter tuning and the training procedures used an early stopping procedure. During the early stopping procedure, the validation data was used to obtain overall validation loss. The minimum required validation loss change was kept at 0 for all experiments. Patience was set to 5 due to a rippling validation loss. The model performance results reported in Section~\ref{sec:hyperparam_tuning} are based on the validation dataset. Reporting on the test dataset during hyperparameter tuning causes test set `peeking'. Cross-validation after tuning on the test set was not considered because it does not lead to more findings with respect to hyperparameter tuning given dataset 1 and 2. % and are reported with help of the confusion matrix summary metrics called accuracy, precision, recall, and Cohen Kappa score.

\begin{figure}[ht]
\centering

\subfloat[\label{fig:tuning_dset1_unstacked_500Hz_rep_sampling_freq}]{\includegraphics[trim={5.3cm 9.4cm 6.2cm 6cm},clip,width=\columnwidth]{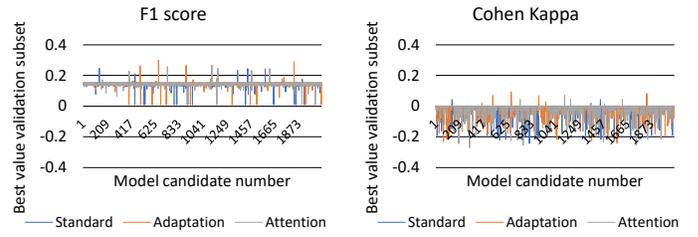}}\hfill
\subfloat[\label{fig:tuning_dset2_stacked_100Hz_rep_sampling_freq}]{\includegraphics[trim={9.0cm 10.1cm 2.5cm 5.4cm},clip,width=\columnwidth]{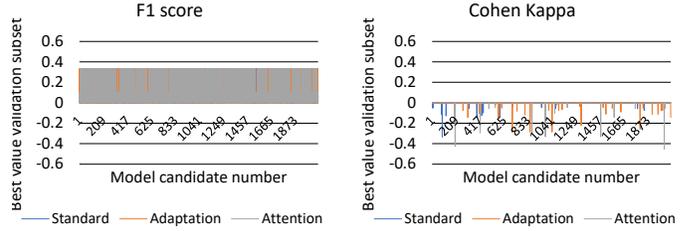}}\hfill
\caption{Hyperparameter tuning results. (a) Dataset 1 of which the samples are diluted via unstacking across RX antenna dimension is considered. The reported sampling frequency to STFT is 500 Hz. (b) Dataset 2 is considered. The reported sampling frequency to STFT is 100 Hz.}
\end{figure}

% \begin{figure}[ht]
%     \centering
%     \includegraphics[trim={4.6cm 5.4cm 6.7cm 4cm},clip,width=\columnwidth]{dset1_unstacked_500Hz_rep_sampling_freq.pdf}
%     \caption{Hyperparameter tuning results. Dataset 1 of which the samples are diluted via unstacking across RX antenna dimension is considered. The reported sampling frequency to STFT is 500 Hz.}
    
% \end{figure}

% \begin{figure}[ht]
%     \centering
%     \includegraphics[trim={4.5cm 5.8cm 6.7cm 4cm},clip,width=\columnwidth]{dset2_stacked_100Hz_rep_sampling_freq.pdf}
%     \caption{Hyperparameter tuning results. Dataset 2 is considered. The reported sampling frequency to STFT is 100 Hz.}
%     \label{fig:tuning_dset2_stacked_100Hz_rep_sampling_freq}
% \end{figure}

In figures~\ref{fig:tuning_dset1_unstacked_500Hz_rep_sampling_freq},~\ref{fig:tuning_dset2_stacked_100Hz_rep_sampling_freq},~\ref{fig:tuning_dset2_unstacked_100Hz_rep_sampling_freq}, and~\ref{fig:tuning_dset2_unstacked_800Hz_rep_sampling_freq}, Cohen Kappa results are either negative or close to zero. This indicates that results do not rise above a random guessing baseline. In figures~\ref{fig:tuning_dset2_unstacked_100Hz_rep_sampling_freq} and~\ref{fig:tuning_dset2_unstacked_800Hz_rep_sampling_freq}, one can see a better Cohen Kappa result for an increased sampling rate communicated to the STFT procedure during pre-processing for the domain adaptation model.

\begin{figure}[ht]
\centering

\subfloat[\label{fig:tuning_dset2_unstacked_100Hz_rep_sampling_freq}]{\includegraphics[trim={8.9cm 10.7cm 2.2cm 5.4cm},clip,width=\columnwidth]{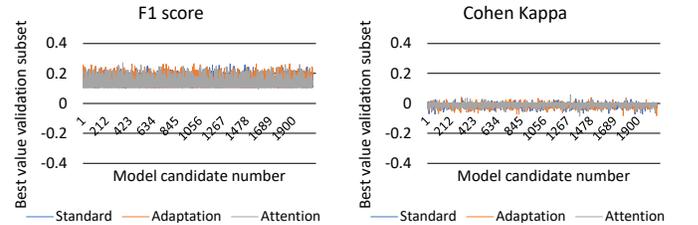}}\hfill
\subfloat[\label{fig:tuning_dset2_unstacked_800Hz_rep_sampling_freq}]{\includegraphics[trim={5.4cm 9.4cm 6.0cm 6.7cm},clip,width=\columnwidth]{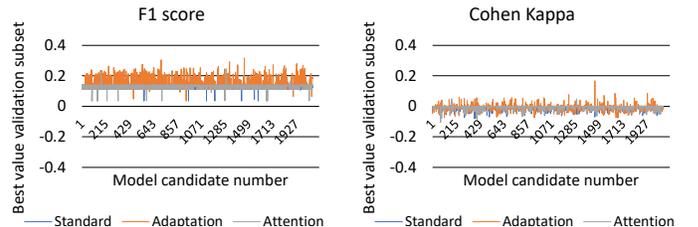}}\hfill
\caption{Hyperparameter tuning results. (a) Dataset 2 of which the samples are diluted via unstacking across RX antenna dimension is considered. The reported sampling frequency to STFT is 100 Hz. (b) Dataset 2 of which the samples are diluted via unstacking across RX antenna dimension is considered. The reported sampling frequency to STFT is 800 Hz.}
\end{figure}

% \begin{figure}[ht]
%     \centering
%     \includegraphics[trim={5.6cm 5.8cm 5.7cm 4cm},clip,width=\columnwidth]{dset2_unstacked_100Hz_rep_sampling_freq.pdf}
%     \caption{Hyperparameter tuning results. Dataset 2 of which the samples are diluted via unstacking across RX antenna dimension is considered. The reported sampling frequency to STFT is 100 Hz.}
%     \label{fig:tuning_dset2_unstacked_100Hz_rep_sampling_freq}
% \end{figure}

% \begin{figure}[ht]
%     \centering
%     \includegraphics[trim={3.8cm 6cm 7.5cm 3cm},clip,width=\columnwidth]{dset2_unstacked_800Hz_rep_sampling_freq.pdf}
%     \caption{Hyperparameter tuning results. Dataset 2 of which the samples are diluted via unstacking across RX antenna dimension is considered. The reported sampling frequency to STFT is 800 Hz.}
%     \label{fig:tuning_dset2_unstacked_800Hz_rep_sampling_freq}
% \end{figure}

\subsection{Hypothesis Discussion}
The DFS input data used for the motion recognition task experiments shows structural differences across Tx beams. Hypothetically, when including all potential structural input differences caused by different Tx beams, of which a single example is given in Figure~\ref{fig:tuning_dset1_unstacked_500Hz_rep_sampling_freq}, one can expect during model inferencing in the training dataset, a DL model could be trained to classify irrespective of these differences. Practically, this will never be the case. When placing a DL model in a setting that is not present in the training dataset, Tx beams, and their interaction with the environment, change thereby causing structural input differences which lead to input representation distribution shifts. These shifts in turn cause performance deterioration. Throughout figures~\ref{fig:vis_inspec_dset1_txb_comparison}, \ref{fig:vis_inspec_dset1_txb9_activity_comparison}, and~\ref{fig:vis_inspec_dset2_txb8_rep_sample_rate_comparison}, many DFS input samples look exactly the same. This means that if input samples are diluted in the Rx antenna dimension, the model will quickly overfit due to confusion over which class these similar images belong to. Therefore, sample dilution in order to increase overall dataset volume is undesirable. Removing specific samples from a dataset because they seem to be always similar irrespective of class, is also not a good solution. The Rx antenna, in which no differences happen is not always the same. This Rx antenna may change during deployment, by switching Tx beam direction, etc. Comparing figures~\ref{fig:vis_inspec_dset1_txb9_activity_comparison} and~\ref{fig:vis_inspec_dset2_txb8_rep_sample_rate_comparison}, many more similarities exist in Figure~\ref{fig:vis_inspec_dset2_txb8_rep_sample_rate_comparison}. This indicates that the 26 GHz cellular multi-user communication device with hybrid beamforming does not have enough sensitivity for allowing one motion per sample repetition. We recommend to increase the number of motions per sample repetition or to reduce the actual sampling rate used during CSI matrix extraction. Reducing the sampling rate results in sharper changes in the time dimension.

Cohen Kappa metrics not rising above a random guessing baseline indicates that a small volume of data is not enough to produce reasonable inference performance. The better Cohen Kappa result for an increased sampling rate communicated to the STFT procedure during pre-processing for the domain adaptation model is likely caused by the fact that higher communicated sampling rates smoothen the DFS input structure. This leads to less task confusion. Sample dilution in the RX antenna dimension overshadows this observation. Therefore, the effectiveness remains inconclusive.

%% Together
\section{Conclusion and Future Work}
In this paper, we introduced human motion recognition as a surrogate task for exploring feasibility of recognizing blockage type using mmWave beamforming data. As recognition model, we used deep learning models capable of dealing with domain factors effects either through extracting domain independent features or domain adaptation. Hypothesis validation has resulted in several recommendations with respect to both DL models and data collection configurations. Presence of domain shift requires use of DL models that can handle it. DFS input data dilution to increase dataset volume should not be considered. A small volume of input data is not enough for reasonable inference performance. Higher sensing resolution, causing lower sensitivity, should be handled by doing more activities/gestures per frame and lowering sampling rate. A higher reported sampling rate to STFT during pre-processing may increase performance, but should always be tested on a per learning task basis.

Our future work directions are four-fold: (i) collecting larger datasets containing more domain factors, (ii) performing a comparison between model-based and data-driven approaches to detect communication blockages based on human motion as a surrogate, (iii) reducing low SNR likelihood by making DL based signal adjustments, and (iv) integrating blockage type recognition into a blockage avoidance pipeline.

\section*{Acknowledgments}
This work has received funding from the European Union’s framework programme for research and innovation Horizon 2020, more specifically MSCA-ITN-2019 MINTS and MSCA-IF-2020 V.I.P., under grant agreement nos. 861222 and 101026885 and the Dutch sectorplan beta en techniek.


\begin{thebibliography}{00}

\bibitem{6G}
C. De Lima et al., "Convergent Communication, Sensing and Localization in 6G Systems: An Overview of Technologies, Opportunities and Challenges," in IEEE Access, vol. 9, 2021%, doi: 10.1109/ACCESS.2021.3053486.

%\bibitem{mmwavenanocell1}
%H. Elshaer, M. N. Kulkarni, F. Boccardi, J. G. Andrews and M. Dohler, "Downlink and Uplink Cell Association With Traditional Macrocells and Millimeter Wave Small Cells," in IEEE Transactions on Wireless Communications, vol. 15, no. 9, 2016%, doi: 10.1109/TWC.2016.2582152.

%\bibitem{mmwavenanocell2}
%H. Q. Ngo, A. Ashikhmin, H. Yang, E. G. Larsson and T. L. Marzetta, "Correction to “Cell-Free Massive MIMO Versus Small Cells” [Mar 17 1834-1850]," in IEEE Transactions on Wireless Communications, vol. 19, no. 5, 2020%, doi: 10.1109/TWC.2020.2974209.

\bibitem{blockage1}
M. Gapeyenko et al., "On the Temporal Effects of Mobile Blockers in Urban Millimeter-Wave Cellular Scenarios," in IEEE Transactions on Vehicular Technology, vol. 66, no. 11, 2017%, doi: 10.1109/TVT.2017.2754543.

%\bibitem{blockage2}
%T. Bai, R. Vaze and R. W. Heath, "Analysis of Blockage Effects on Urban Cellular Networks," in IEEE Transactions on Wireless Communications, vol. 13, no. 9, 2014%, doi: 10.1109/TWC.2014.2331971.

\bibitem{camera}
Y. Oguma, T. Nishio, K. Yamamoto \emph{et~al.}, "Performance modeling of camera-assisted proactive base station selection for human blockage problem in mmWave communications", IEEE WCNC, 2016%, pp. 1-7, doi: 10.1109/WCNC.2016.7564769.

%\bibitem{camera2}
%T. Nishio et al., "Proactive Received Power Prediction Using Machine Learning and Depth Images for mmWave Networks," in IEEE Journal on Selected Areas in Communications, vol. 37, no. 11, 2019%, doi: 10.1109/JSAC.2019.2933763.

%\bibitem{radar}
%U. Demirhan and A. Alkhateeb, “Radar aided proactive blockage prediction in real-world millimeter wave systems,” arXiv preprint arXiv:2111.14805, 2021

%\bibitem{lidar}
%D. Marasinghe, N. Rajatheva, and M. Latva-aho, “LiDAR Aided Human Blockage Prediction for 6G,” in IEEE Globecom Workshops (GCWkshps), 2021%, pp. 1–6.

\bibitem{sub6}
M. Alrabeiah and A. Alkhateeb, "Deep Learning for mmWave Beam and Blockage Prediction Using Sub-6 GHz Channels," in IEEE Transactions on Communications, vol. 68, no. 9, 2020%, doi: 10.1109/TCOMM.2020.3003670.

\bibitem{in-bandb}
S. Wu, M. Alrabeiah, C. Chakrabarti, and A. Alkhateeb, “Blockage
prediction using wireless signatures: Deep learning enables real-world
demonstration,” arXiv preprint arXiv:2111.08242, 2021


\bibitem{JCS_mmwave_butler}
T. M. Pham, R. Bomfin \emph{et~al.}, "Joint Communications and Sensing Experiments Using mmWave Platforms," IEEE SPAWC, 2021%, pp. 501-505, doi: 10.1109/SPAWC51858.2021.9593174.

\bibitem{1603402}
V.~Chen, F.~Li, S.-S. Ho, and H.~Wechsler, ``{Micro-Doppler effect in radar:
  phenomenon, model, and simulation study},'' \emph{IEEE Trans. Aerospace and
  Electronic Systems}, vol.~42, no.~1%, pp. 2--21, 2006.

%\bibitem{10.1145/2971648.2971670}
%W.~Wang, A.~X. Liu, and M.~Shahzad, ``{Gait Recognition Using Wifi Signals},''
 % in \emph{Proc. ACM Int. Conf. UbiComp}, 2016%, p. 363–373.

\bibitem{10.1145/3307334.3326081}
Y.~Zheng, Y.~Zhang \emph{et~al.}, ``{Zero-Effort Cross-Domain Gesture
  Recognition with Wi-Fi},'' in \emph{Proc. ACM MobiSys}, 2019%, pp.
  313--325.

\bibitem{10.5555/3016100.3016186}
B.~Sun, J.~Feng, and K.~Saenko, ``{Return of Frustratingly Easy Domain
  Adaptation},'' in \emph{Proc. AAAI Conf.}, 2016%, pp. 2058--2065.

\bibitem{wigrunt_gu_et_al_2021}
Y.~Gu, X.~Zhang, Y.~Wang \emph{et~al.}, ``{WiGRUNT: WiFi-enabled Gesture
  Recognition Using Dual-attention Network},'' 2021, techRxiv Preprint
  %https://doi.org/10.36227/techrxiv.13846052.v1.

\bibitem{8839094}
J.~Yang, H.~Zou, Y.~Zhou \emph{et~al.}, ``{Learning Gestures From WiFi: A
  Siamese Recurrent Convolutional Architecture},'' \emph{IEEE Internet of
  Things J.}, vol.~6, no.~6, 2019

\bibitem{10.1145/3366423.3380091}
X.~Chen, H.~Li, C.~Zhou \emph{et~al.}, ``{FiDo: Ubiquitous Fine-Grained
  WiFi-Based Localization for Unlabelled Users via Domain Adaptation},'' in
  \emph{Proc.ACM Web Conf.}, 2020%, p. 23–33.

\bibitem{10.1145/3408308.3427983}
E.~Soltanaghaei, R.~A. Sharma, Z.~Wang \emph{et~al.}, ``{Robust and Practical
  WiFi Human Sensing Using On-Device Learning with a Domain Adaptive Model},''
  in \emph{Proc. ACM BuildSys}, 2020%, p. 150–159.

\bibitem{10.1145/3241539.3241548}
W.~Jiang, C.~Miao \emph{et~al.}, ``{Towards Environment Independent
  Device Free Human Activity Recognition},'' in \emph{Proc. ACM MobiCom}, 2018%, pp. 289--304.

\bibitem{10.1145/3380980}
H.~Xue, W.~Jiang, C.~Miao \emph{et~al.}, ``{DeepMV: Multi-View Deep Learning
  for Device-Free Human Activity Recognition},'' \emph{Proc. ACM IMWUT},
  vol.~4, no.~1, 2020%, art. 34, 26 pp.

\bibitem{9413590}
Z.~Wang, S.~Chen, W.~Yang \emph{et~al.}, ``{Environment-Independent Wi-Fi Human
  Activity Recognition with Adversarial Network},'' in \emph{Proc. IEEE
  ICASSP}, 2021%, pp. 3330--3334.

%\bibitem{subj_ind_zhou_et_al_2020}
%S.~Zhou, L.~Guo, Z.~Lu \emph{et~al.}, ``{Subject-independent Human Pose Image
 % Construction with Commodity Wi-Fi},'' 2020, arXiv:2012.11812

\bibitem{10.1049/ell2.12097}
Y.~Yin, Z.~Zhang, X.~Yang \emph{et~al.}, ``{Towards fully domain-independent
  gesture recognition using COTS WiFi device},'' \emph{IET Electron. Lett.},
  vol.~57, no.~5, 2021

\bibitem{9488693}
D.~Li, J.~Xu, Z.~Yang \emph{et~al.}, ``{Train Once, Locate Anytime for Anyone:
  Adversarial Learning based Wireless Localization},'' in \emph{Proc. IEEE
  INFOCOM}, 2021%, pp. 1--10.

%\bibitem{9068635}
%F.~Wang, J.~Liu, and W.~Gong, ``{WiCAR: WiFi-based in-Car Activity Recognition
  %with Multi-Adversarial Domain Adaptation},'' in \emph{Proc. IEEE/ACM IWQoS},
 % 2019%, pp. 1--10.

%\bibitem{adversary_helps_liu_et_al_2020}
%J.~Liu, J.~Han, F.~Lin \emph{et~al.}, ``{Adversary Helps: Gradient-based
  %Device-Free Domain-Independent Gesture Recognition},'' 2020, arXiv:2004.03961

\bibitem{9149596}
X.~Yang, R.~Cao, M.~Zhou \emph{et~al.}, ``{Temporal-Frequency Attention-Based
  Human Activity Recognition Using Commercial WiFi Devices},'' \emph{IEEE
  Access}, vol.~8, 2020

%\bibitem{Goodfellow-et-al-2016}
%I.~Goodfellow, Y.~Bengio, and A.~Courville, \emph{{Deep Learning}}.\hskip 1em
 % plus 0.5em minus 0.4em\relax MIT Press, 2016,
 % Available: http://www.deeplearningbook.org.

\bibitem{10.5555/3122009.3242042}
L.~Li, K.~Jamieson, G.~DeSalvo \emph{et~al.}, ``{Hyperband: A Novel
  Bandit-Based Approach to Hyperparameter Optimization},'' \emph{JMLR},
  vol.~18, no.~1, pp. 6765--6816, Jan. 2017.

\bibitem{8578572}
M.~Sandler, A.~Howard, M.~Zhu \emph{et~al.}, ``{MobileNetV2: Inverted Residuals
  and Linear Bottlenecks},'' in \emph{Proc. IEEE/CVF Conf. CVPR}, 2018

\bibitem{8701503}
J.~Hu, L.~Shen, S.~Albanie \emph{et~al.}, ``{Squeeze-and-Excitation
  Networks},'' \emph{IEEE TPAMI}, vol.~42, no.~8, 2020

\end{thebibliography}
\end{document}